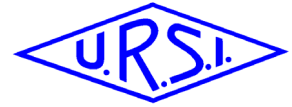

# Impact of new 5G network components on out-of-band emissions at 23.8 GHz

Elliot Eichen
Wireless Interdisciplinary Research Group
Computer Science Department
University of Colorado
Boulder, CO 80309, USA

## Abstract

The impact of 5G networks transmitting between 24.25-27.5 GHz on Earth Exploration Satellite Services (EESS) microwave sounders used to measure atmospheric water vapor and temperature was widely discussed and modeled in preparation for setting emission recommendations by International Telecommunications Union (ITU) at the 2019 World Radio Congress (WRC-19). Since then, two new classes of network devices – 5G repeaters and high transmission power User Equipment (UE) for fixed-wireless services – have been introduced and deployed in 28 GHz networks with expectations that they will also be deployed at 24 GHz. This paper discusses the (potentially significant) increase in interference from these new components along with open questions related to their regulatory status. While this paper discusses increases in interference to 23.8 GHz EESS measurements from 5G transmissions in the "24 GHz" band, *it is important to recognize that repeaters and high power UEs need to be considered when modeling interference from 5G/6G networks in all bands*. This paper also touches on whether the current ITU process and methodology to regulate interference with passive sensors (vendor applied hardware-based filtering based on long-term network forecasts and worst-case Monte Carlo modeling) can keep up with rapidly changing wireless technology and the increased competition for spectrum.

## 1 Introduction

The impact of 5G networks transmitting at 24.25-24.45 GHz (3GPP n258 a&b blocks – figure 1) on microwave sounders used to measure atmospheric water vapor and temperature was widely discussed and modeled in preparation for setting emission recommendations at ITU's WRC-19 [1][2][3]. Since then, two new classes of network devices – 5G repeaters [4][5] and 3GPP Power Class 1 User Equipment [6][7] – have been introduced and deployed in 28 GHz networks with expectations that these components will also be deployed in other mm and sub-mm wave bands including 24 GHz. This paper discusses the potentially significant increase in interference (compared with estimates presented at WRC-19) from these new components and open questions about their regulatory status.

In the case of 5G repeaters, an increase in Out-of-Band Emissions (OOBE) is due to an increase in the density of downstream (base-station to UE), upstream (UE to base station), and base-station to base-station transmitters. The salient questions are how much additional interference is added and whether it can be mitigated. From a regulatory perspective, there is also uncertainty in whether WRC-19 recommendations adopted to limit interference from base stations (gNBs) and User Equipment (UEs) apply to repeaters[1].

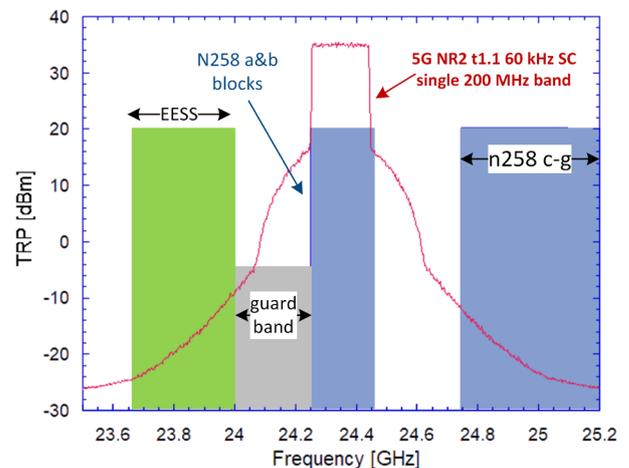

**Figure 1**. Allocated frequency bands for EESS microwave sounders (green), 5G @ 24 GHz (blue), a guard band separating EESS from 5G (gray), with a 5G test spectrum for a 5G repeater (37 dBm) superimposed on top.

The case for increased interference from high power fixed-wireless UEs (3GPP class 1 = 40 dBm) compared to WRC-19 estimates is more subtle than for 5G repeaters. High power UEs <u>will</u> contribute higher aggregate OOBE even while complying with ITU recommendations <u>if</u> they do not implement the 3GPP UE Power Control Algorithm ($P_{PwrCtl}$) [8] designed to decrease battery drain from smartphones. Since fixed-wireless UEs are line powered

(rather than battery-powered), and the performance of fixed-wireless services can be improved by not adopting $P_{PwrCtl}$, it is important to clarify whether or not these UEs implement a power control algorithm[ii].

In addition to concern over $P_{PwrCtl}$ implementations, there is also concern around whether high power fixed-wireless UEs will be exempted from WRC-19 recommendations. High transmission power UE compliance with WRC-19 UE OOB emission recommendations is more difficult than the equivalent compliance for gNBs because the price of UEs (customer premise equipment) is significantly less elastic than the price of a gNB (a gNB is between two and three orders of magnitude more expensive than a 5G UE modem). It is thus not surprising that at least one vendor of high-power fixed-wireless chipsets has petitioned the US Federal Communications Commission (FCC) for exemption from compliance with ITU recommendations [9].

## 2 5G Repeaters

Access repeaters (gNB⇔UEs) were introduced and deployed starting in 4Q2019 as a cost-effective [iii] mechanism to improve UE coverage within 5G 28GHz cells. Similar repeaters have also been deployed at 28 and 39 GHz to create RF mesh networks by linking gNBs. These networks (typically referred to as Integrated Access Backhaul or IAB networks [10][16]) use repeaters to simultaneously support trunking (gNB⇔gNB) and access (gNB⇔UEs). They reduce the cost of 5G/6G networks by lowering the number of optical fiber interconnects to the 5G/6G core. This paper refers to these components as IAB repeaters.

While access repeaters and IAB repeaters are currently being deployed at 28 GHz, these devices are expected to be deployed in other mm and sub-mm wave bands, including 24 GHz. (Note that other network components - typically customer premise equipment - that relay 5G transmissions from an outside macro wireless network to the interior of residences and commercial buildings[iv] replace rather than add UE transmitters. Therefore, they are not considered here, although they may play a role in increased OOBE as suggested in the next section on 3GPP power class 1 devices.)

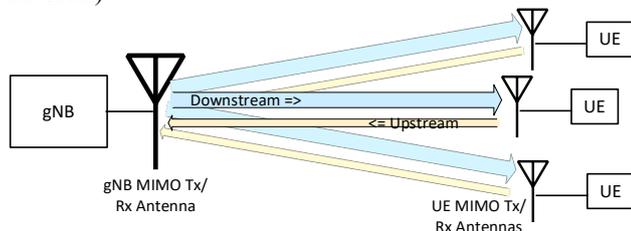

**Figure 2a.** 5G network block diagram without repeaters. MIMO capability provides spatial multiplexing and gain. Although not shown figuratively, mm and sub-mm wave 5G/6G transmissions use line-of-sight (LOS) and non-line-of-sight (NLOS) paths

Block diagrams for networks with and without access repeaters and with IAB repeaters are shown separately in figures 2a, 2b, and 2c. Actual deployments will have various (and probably dynamic) combinations of direct gNB to UE connections and gNB to UE connections mediated by a repeater.

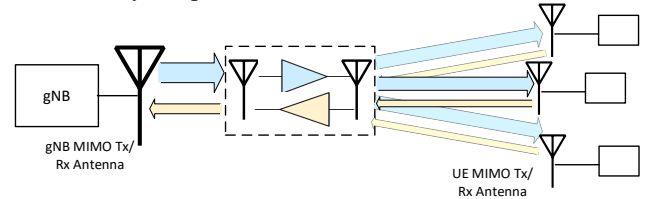

**Figure 2b:** 5G network block diagram with an access repeater.

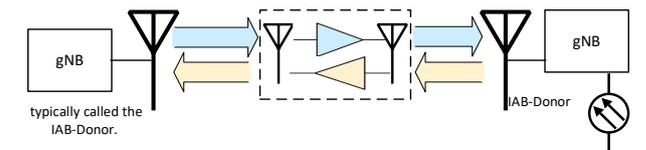

**Figure 2c:** 5G/6G IAB repeater shown to link gNBs. IAB architectures would support a combination of figure 2b and 2c links. While this figure shows two gNBs sharing a single fiber connection to the core network, IAB links can be deployed in a mesh, supporting N gNBs with M fiber connections with N > M.

While it is clear that 5G/6G repeaters will increase OOBE (more transmitters in a given cell => more OOBE), the salient question is how much additional OOBE is created, how does that translate to interference with Earth Exploration Satellite Services (EESS) microwave sounders, and what is the appropriate way to mitigate that interference. The ITU provides a recipe for modeling OOBE and interference based on a Monte Carlo network simulation and a complex set of assumptions for network infrastructure, RF propagation, and microwave sounder properties [11]. Extending the ITU's model to include repeaters requires adding realistic assumptions for repeater density, location, and operating parameters; however, adding realistic repeater assumptions may be challenging, particularly in the short term as network architectures and deployments continue to evolve.

Prior to adopting assumptions necessary to model the impact of repeaters, it is possible to approximate the impact that access repeaters may have on interference based on the assumption that each repeater will contribute roughly the same OOBE as a gNB downstream or the same number of subtended UEs upstream. There is some justification for this approximation if (as suggested by repeater vendors) repeaters are regulated as "Class B industrial signal boosters" [12][v]. In this case, the FCC limits OOBE to -70dBm integrated over a 10 kHz band; this is equivalent to -27 dBm integrated over 200MHz, which is close to the OOBE limits for gNBs and UEs adopted at WRC-19[1]. This forms the justification for approximating OOBE from one repeater as adding one gNB in the downstream direction, and doubling the number of subtended UEs in the upstream direction.

Following [11] and [13] the RF power $P_j^{sat}$ measured within a specified frequency bandwidth by a microwave sounder from any j$_{th}$ transmission antenna is:

$$P_j^{sat} = \frac{P_j^{Tx} Tx_j^{gain}(\theta_{gNB}, \Phi_{gNB}) Sat^{gain}(\theta_{EESS}, \Phi_{EESS})}{L_{path} L_{gas} L_{other}} \quad (1)$$

where $P_j^{Tx}$ is the total radiated power within the specified bandwidth (TRP), $Tx_j^{gain}(\theta_{gNB}, \Phi_{gNB})$ is the gain of the j$_{th}$ transmitting antenna gain as a function of elevation and azimuth, $Sat^{gain}(\theta_{EESS}, \Phi_{EESS})$ is the sounder antenna gain as a function of elevation and azimuth with respect to the sounder sensor axis, $L_{path}$ is the ($1/r^2$) free-space path loss, $L_{gas}$ is the attenuation due to gaseous absorption, and $L_{other}$ includes polarization, clutter, and other potential fixed losses. Equation (1) holds for all transmitting antennas; Tx can be a gNB, a repeater (Rep), or UE. The only difference is that UEs in the ITU model includes an additional term ($L_{PwrCtl}$) which comes from a 3GPP specification to decrease battery drain from smartphones by reducing UE transmit power while maintaining an acceptable carrier-to-noise ratio.

The total RF power measured by the microwave sounder is the sum over all $j$ transmitters. The total power measured by the sounder in the downstream direction ($P^{sat-down}$) is thus:

$$P^{sat-down} = \sum_j^{N_{gNB}} P_j^{sat-gNB} + \sum_k^{N_{Rep}} P_k^{sat-Rep} \quad (2)$$

where $N_{gNB}$ is the number of gNBs subtended by the effective field of view of the sounder (the measurement pixel) and $N_{rep}$ is the number of repeaters subtended by the effective field of view of the sounder.

If $N_{Rep} = F\, N_{gNB}$ (i.e., there are F repeaters in each gNB cell), and $<P^{sat-Rep}> \approx F <P^{sat-gNB}>$ (the average OOBE emission power from each repeater is approximately the same as the average OOBE power from each gNB), then the total downstream OOBE becomes

$$P^{sat-down} \approx (1+F)\, P^{sat-gNB} \quad (3)$$

The approximate downstream interference penalty in dB for a network with repeaters, compared with the same network without repeaters, is thus $10 \log(1+F)\,[dBi]$. A similar equation exists for the interference penalty in the upstream direction.

How many repeaters are there likely to be (particularly in urban hotspot areas which are most likely to interfere with sounder measurements) within a given cell? This data is hard to come by as network planners for 5G operators have declined to provide estimates. Although some repeater vendors have suggested that there could be as many as 8-10 repeaters (as a combination of access and IAB devices) per cell, it seems likely to the author [vi] that F will be between 1 and 4, thus yielding an additional interference penalty of approximately 3 to 7 dBi. (Note that 3GPP rel 18 may contain provisions for duplex transmission [14]. If duplex transmission is used in a given deployment, the interference penalty would then be the sum of both upstream and downstream penalties.)

### 3 High Power Fixed-Wireless User Equipment

The reason for a potential increase in interference from high power fixed-wireless UEs (3GPP class 1 = 40 dBm EIRP) [6] compared to UE assumptions in WRC-19 estimates is more subtle than for 5G repeaters. UEs are generally considered to be battery-powered mobile devices[vii]. Power management of these devices to extend their operating time before recharging is a primary concern for the wireless community. The 3GPP UE Power Control Algorithm (P$_{PwrCtl}$)[8] reduces battery drain by lowering the UEs transmission power below the maximum allowed power as long as the Carrier-to-Noise ratio detected by the gNB is acceptable.

Fixed-Wireless devices however are not battery powered and thus do not benefit from implementing P$_{PwrCtl}$. Moreover, performance degradation due to 5G channel fading can be reduced by not implementing this capability. The author has repeatedly asked fixed wireless UE vendors, and also the underlying chip-sets manufacturers, whether the P$_{PwrCtl}$ is implemented in these devices but has not received any response.

Since estimates of interference to EESS measurements at 23.8GHz for WRC-19 assume P$_{PwrCtl}$, it is interesting to understand the impact if this capability is not implemented. As stated earlier, equation (1) for the power received at the microwave sounder from the j$_{th}$ UE includes $L_{PwrCtl}$ (the additional reduction in received power due to P$_{PwrCtl}$) and is given by:

$$P_j^{sat} = \frac{P_j^{Tx-UE} Tx_j^{gain}(\theta_{gNB}, \Phi_{gNB}) Sat^{gain}(\theta_{EESS}, \Phi_{EESS})}{L_{PwrCtl} L_{path} L_{gas} L_{other}} \quad (4)$$

A Monte Carlo simulation without repeaters based on ITU recommended assumptions (and with a UE maximum TRP of 22 dBm) concluded that $L_{PwrCtl}$ reduced $P_j^{sat}$ by 0 to 63 dB [13]. The OOBE inference limits from this study were consistent with other submissions to ITU committee 5/1 considering interference from 24 GHz 5G [15]. **It is thus not hard to conclude that a UE without P$_{PwrCtl}$ (and with a EIRP > 40dBm) would significantly increase the RF power from 5G UEs detected by microwave sounders. Moreover, this increase in detected RF would occur even if the UE was in compliance with ITU recommended OOBE emission limits.**

To quantify the increase in interference, the author has suggested to the institutional (government) owners of the

US interference model that it would be interesting to run their simulation without the $P_{PwrCtl}$ term. However, there has been no response to date.

In addition to the increase in OOBE emissions related to $P_{PwrCtl}$, there is also a question related to the economic viability of complying with WRC-19 recommendations for high power fixed-wireless UEs. Network equipment vendors have expressed concern over meeting gNB WRC-19 OOBE recommendations (particularly the more stringent limits that begin in 2027). Given that the price of a UE (particularly those that compete with traditional wireline broadband customer premise equipment) is constrained, it is not surprising that chipset vendors for high power devices have requested an exemption for WRC-19 recommendations based on economic constraints [9].

## 4 Is the ITU Process and Model Outdated?

The introduction and deployment of new components to the 5G ecosystem before they can be considered by national regulators or included in the ITU's process to protect EESS resources is, in this author's opinion, symptomatic of a regulatory infrastructure that has not evolved to meet present-day challenges. The wireless ecosystem moves at internet speed while ITU (and many national regulatory) processes were built for the PSTN era. The slow pace of the regulatory cycle creates uncertainty in the deployment of new capabilities and can lead to un-intended degradation of existing service and legacy applications.

While a complete analysis and discussion of the regulatory apparatus is beyond the scope of this paper, suggestions for improvement might include:

- Providing national regulatory agencies and the ITU with the technology leadership and engineering resources required to keep pace with the speed of innovation, development, and deployment in the wireless arena.

- Implementing an expedited process (perhaps triggered by a small but reasonable number of member counties) to consider new network architectures and devices compared to the ITU's four-to-eight-year cycle time for regulatory oversight.

- Consider alternate methodologies that prevent interference with EESS passive sensors and simultaneously encourage spectrum sharing. The existing methodology – vendor applied hardware-based frequency filtering based on long term network forecasts and worst-case Monte Carlo modeling – could be supplemented by software-based methods [17][18] that are less expensive, faster to adapt to network changes, and are equal to or potentially better at preventing interference.

## 5 Summary

This paper discussed the increase in interference to EESS microwave sounds at 2.38GHz from new 5G/6G network components – repeaters (for both UE access and backhaul), and high power (3GPP Class 1) fixed-wireless UEs. These components were introduced after OOBE emission recommendations were agreed upon at WRC-19. These components have been deployed in the "28 GHz" band and are available and expected to be deployed in the "24 GHz" 5G band and other mm and sub-mm bands.

The approximate OOBE penalty due 5G repeaters for gNBs is given by ~10log(1+F), where F is the number of access pus IAB repeaters/cell, with a similar expression for UEs. The additional penalty is due to the increase in the density of transmission antennas per cell above the density assumed in ITU recommendations. For reasonable values of F (2-4), the penalty is ~3-7 dBi. Since repeaters were not included in estimates leading to WRC-19 recommendations but will have an impact on OOBE, these recommendations should be updated to account for the presence of these devices.

A potentially substantive increase in interference from high power fixed-wireless UEs could occur if these devices do not implement a 3GPP standard designed to lower UE batter drain. There is reason to believe that UEs do not support this capability as these devices are line (rather than battery) powered. Moreover, the performance of 5G networks would benefit from not implementing $P_{PwrCtl}$ due to a reduction in channel fading. It is therefore imperative for UE vendors to clarify whether their devices implement $P_{PwrCtl}$.

It is not possible to quantitatively access the impact of turning off $P_{PwrCtl}$ without performing a Monte-Carlo simulation using ITU assumptions. However, one simulation that included $P_{PwrCtl}$ concluded that this capability reduced the RF received by the microwave sensor from subtending UEs by between 0 and 63 dB. It is thus not hard to imagine that a fixed-wireless deployment of UEs without $P_{PwrCtl}$ could significantly increase interference with microwave sounder measurements. Moreover, such an increase in interference would occur even if the UE complied with the ITU recommended OOBE emission limits agreed upon at WRC-19. It would be beneficial (and presumably easy) for one of the national entities that provided estimates of interference to the 23.8 GHz band to run their simulation without $P_{PwrCtl}$ to quantify this impact.

---

[i] A US manufacturer of 5G repeaters has told the author that 5G repeaters are regulated as Industrial Signal Boosters. The author has requested clarification from the FCC re the regulatory status of these devices but has not received a reply.

[ii] The author has repeatedly asked manufacturers of high power fixed-wireless UEs and manufacturers of the chipsets used in these devices whether $P_{PwrCtl}$ is implemented or whether it can be disabled, but has not received a reply.

[iii] The cost of a 5G network repeater is estimated to be between 10 and 50% the cost of a 5G base station (gNB).

[iv] For example, antennas that sit outside buildings and relay signals optically through a window or with coax to an internal antenna. Other examples include Distributed Antenna Systems (DAS) which typically combine coax-based transport and signal processing to relay the wireless WAN to the inside of commercial buildings.

[v] The author has asked the FCC for clarification on whether 5G/6G repeaters are regulated as class B industrial signal boosters, and whether they should inherit the same WRC-19 recommendations as gNBs and UEs, but has not received a reply.

[vi] Based on private informal conversations with equipment suppliers and network engineers.

[vii] While there are some wireless devices (such as VoLTE desk phones, or 4G connected routers) that are not battery powered, the deployed number of these devices is miniscule compared to the number of mobile phones and smartphones. Thus, there has not been much discussion of UE's not implementing this capability.